\documentclass[review]{elsarticle}
\usepackage[MeX,plmath]{polski}
\usepackage[utf8]{inputenc}
\usepackage[polish,british,UKenglish,USenglish,english,american]{babel}
\usepackage{graphicx}
\usepackage{amsmath}
\usepackage{braket}
\usepackage[hidelinks]{hyperref}

\journal{Nukleonika}

\begin{document}

\begin{frontmatter}
\title{Searches for discrete symmetries violation in ortho-positronium decay using the J-PET detector}

\author[WFAIS]{D.~Kamińska}
\author[WFAIS]{T.~Bednarski} 
\author[WFAIS]{P.~Białas} 
\author[WFAIS]{E.~Czerwiński} 
\author[WFAIS]{A.~Gajos} 
\author[UMCS]{M.~Gorgol} 
\author[UMCS]{B.~Jasińska}
\author[WFAIS,PAN]{Ł.~Kapłon}
\author[WFAIS]{G.~Korcyl} 
\author[SWIERK]{P.~Kowalski} 
\author[WFAIS]{T.~Kozik}
\author[SWIERKHEP]{W.~Krzemień} 
\author[WFAIS]{E.~Kubicz} 
\author[WFAIS]{Sz.~Niedźwiecki} 
\author[WFAIS]{M.~Pałka} 
\author[SWIERK]{L.~Raczyński} 
\author[WFAIS]{Z.~Rudy} 
\author[WFAIS]{O.~Rundel}
\author[WFAIS]{N.G.~Sharma}
\author[WFAIS]{M.~Silarski} 
\author[WFAIS]{A.~Słomski}
\author[WFAIS]{A.~Strzelecki} 
\author[WFAIS,PAN]{A.~Wieczorek} 
\author[SWIERK]{W.~Wiślicki} 
\author[WFAIS]{M.~Zieliński} 
\author[UMCS]{B.~Zgardzińska} 
\author[WFAIS]{P.~Moskal}
%
%
\address[WFAIS]{Faculty of Physics, Astronomy and Applied Computer Science, Jagiellonian University, \newline S.~Łojasiewicza 11, 30-348 Kraków, Poland}
\address[UMCS]{Department of Nuclear Methods, Institute of Physics, Maria Curie-Sklodowska University, \newline Pl. M. Curie-Sklodowskiej 1, 20-031 Lublin, Poland}
\address[PAN]{Institute of Metallurgy and Materials Science of Polish Academy of Sciences, \newline W. Reymonta 25, 30-059 Kraków, Poland}
\address[SWIERK]{Świerk Computing Centre, National Centre for Nuclear Research, \newline  05-400 Otwock-Świerk, Poland}
\address[SWIERKHEP]{High Energy Department, National Centre for Nuclear Research, \newline 05-400 Otwock-Świerk, Poland}

%
\begin{abstract}

In this paper we present prospects for using the J-PET  detector  to  search
for discrete symmetries violations  in  a  purely  leptonic  system  of  the
positronium atom. We discuss tests of $\mathcal{CP}$ and  
$\mathcal{CPT}$  symmetries  by  means  of
ortho-positronium decays into three photons. No zero expectation values  for
chosen correlations between ortho-positronium spin and momentum  vectors  of
photons would imply the existence of physics phenomena beyond  the  Standard
Model. Previous measurements resulted in violation amplitude parameters  for
$\mathcal{CP}$ and $\mathcal{CPT}$ symmetries consistent with zero, with an uncertainty of about 
$10^{-3}$.  The  J-PET  detector  allows  to  determine  those  values  with  better
precision thanks to a unique time and angular  resolution  combined  with  a
high geometrical acceptance. Achieving the aforementioned  is  possible  due
to application of polymer scintillators instead of crystals as detectors  of
annihilation quanta.
\end{abstract}

\begin{keyword}
 \texttt{discrete symmetries} \sep \texttt{J-PET} \sep \texttt{ortho-positronium}
\end{keyword}
\end{frontmatter}

\section{Introduction}

One of the important issues of physics nowadays is  validation  of  discrete
symmetries: charge conjugation ($\mathcal{C}$), space reflection ($\mathcal{P}$) and  
time  reversal~($\mathcal{T}$), and combinations of them. These  problems,  extensively  studied  since
decades in elementary processes governed by electroweak forces,  still  need
to be measured with higher accuracy in order  to  explain  such  fundamental
questions as: predominance of matter over  antimatter  in  the  Universe  or
validity of the Lorentz invariance.
In  those  studies  the  lightest  strange  meson  sector  occurs  specially
fruitful. The violation of $\mathcal{CP}$ and $\mathcal{T}$ symmetries were observed  by  J.~Cronin
and V.~Fitch~\cite{Christenson:1964fg} and  by  the  BABAR  Collaboration~\cite{Aubert:2001sp}. 
Surprisingly,  in
lepton sector there is no  indication  of  the  $\mathcal{T}$,  $\mathcal{CP}$  and  $\mathcal{CPT}$  symmetries
violation. It is important to emphasize that the presently known sources  of
the $\mathcal{CP}$ symmetry violations are still too small to account for  the  observed
excess of matter over antimatter~\cite{Sakharov:1967dj} and this remains one  of  the  greatest
puzzles in physics and cosmology.
In this  paper  we  focus  on  $\mathcal{CP}$  and  $\mathcal{CPT}$  symmetry  tests  in  decays  of
positronium and perspectives of their investigation by means  of  the  J-PET detector.

\section{Violation of discrete symmetries in ortho-positronium decays}

A special role in discrete symmetry violation searches plays  a  positronium
atom, which due to its sensitivity~\cite{Bernreuther:1988tt} to a variety  of  symmetry  violation
effects, is one of the best candidates for such kind of studies.
The  positronium  atom  structure  is  analogous  to  the  Bohr  atom.   The
ortho-positronium triplet (o-Ps) and para-positronium  singlet  (p-Ps)  states
can be distinguished and their spin alignment determines  their  properties.
Due to the charge conjugation conservation the o-Ps can decay only into  odd
number of photons, while the p-Ps decays into even number  of  photons,  and
the mean lifetime of o-Ps state in vacuum is longer (140 ns~\cite{Harpen:2003zz}) than for p-Ps
 state (120 ps~\cite{Harpen:2003zz}).

Studies of discrete symmetries  violation  in  ortho-positronium  state  were
proposed by Bernreuther~et~al. in  1988~\cite{Bernreuther:1988tt}.  The  signals  for  discrete
symmetries violation in a spin-polarized ortho-positronium  will  be  visible
in selected set of angular correlations build on $i$-th photon  momentum  $\vec{k}_i$
(photons are numbered in order of decreasing  energy)  and  ortho-positronium
spin $\vec{S}$. The evidence for discrete symmetry violations  will  be  observed
in non-vanishing value of one of forbidden correlations 
(e.g. $\vec{S} \cdot \hat{k}_1 \times \hat{k}_2$ for  $\mathcal{CPT}$
symmetry).

The measured observable is the asymmetry:
\begin{equation}
    A = \frac{N_+ - N_-}{N_+ + N_-},
\end{equation}
where $N_+$ and $N_-$ denotes number of decays with the normal to the decay
plane parallel (+) and antiparallel (-) to the spin direction,
respectively.
Asymmetry value can be associated to the $\mathcal{CP}$ ($C_{CP}$) and $\mathcal{CPT}$ ($C_{CPT}$)
violation parameters by equations:
\begin{align}
    A  = C_{CP} \cdot S^{CP}, \\
    A  = C_{CPT} \cdot S^{CPT},
\end{align}
where $S^{CP}$ and $S^{CPT}$ are the analyzing powers build on operators  
$( \hat{S} \cdot \hat{k}_1 )  (\hat{S} \cdot \hat{k}_1 \times \hat{k}_2 )$
and $\vec{S} \cdot \hat{k}_1 \times \hat{k}_2$, respectively.

\section{Experimental verification}
\subsection{$\mathcal{CP}$ symmetry}
The most recent measurement was presented at Tokyo University  in  2010~\cite{Yamazaki:2009hp}.
Positrons emitted from the 1 MBq $^{22}$Na source at the center  of  experimental
setup were passing through plastic scintillators and  bound  with  electrons
in silica aerogel inserted in the external 5 kG magnetic field.  The  gamma-
rays emitted from oPs decay were registered by LYSO crystals.
The measured value of  $\mathcal{CP}$ violating parameter is equal to~\cite{Yamazaki:2009hp}:
\begin{equation}
    C_{CP} = 0.0013 \pm 0.0012_{\mbox{stat}} \pm 0.0006_{\mbox{syst}}.
\end{equation}
Precision of obtained result  is  limited  by  available  statistics,  which
cannot be increased by higher intensities  of  radioactive  sources  due  to
pile-ups in detector system~\cite{Yamazaki:2009hp}.

\subsection{$\mathcal{CPT}$ symmetry}
The $\mathcal{CPT}$ violation coefficient was measured by Vetter and Freedman using  the
Gammasphere detector~\cite{vetter:2003}  -  a  [pic]  spectrometer  for  nuclear  structure
research  built  by  110  high-purity  germanium   detectors.   During   the
experiment the $^{68}$Ge and $^{22}$Na positron sources  were  used,  with  quite  low
intensities 0.04MBq to avoid  pile-ups  in  detector.  Ortho-positronium  was
formed in silicon dioxide aerogel and decays into  three  gammas  that  were
registered  by  detector.  Reconstruction  of  [pic]ortho-positronium  decays
allows to determine the following $\mathcal{CPT}$ violation coefficient~\cite{vetter:2003}:
\begin{equation}
    C_{CPT} = 0.0071 \pm 0.0062.
\end{equation}
Obtained result is the most precise measurement till now.

\section{Prospects for J-PET}

Jagiellonian Positron Emission Tomograph (J-PET)  is  a  detector  based  on
plastic scintillators characterized by  shorter  signals  (about  5ns)  than
commonly used crystal scintillators (e.g.  50  ns  for  GSO  crystal)~\cite{Moskal:2014rja,Moskal:2014sra}.
This allows to use high  intensity  sources  and  fast  digital  electronics
readout~\cite{Korcyl:2014,Palka:2014,Krzemien:2015hkb}. Compton scattering spectrum instead  of  photopeak  can  be
used by applying a dedicated analysis~\cite{Raczynski:2014poa,Raczynski:2015zca,Moskal:2015jzy}.
As a preparation for  this  project,  a  series  of  simulations  have  been
carried out in order to estimate physical and  instrumental  background  for
studies  of  discrete  symmetries.   They   account   for   the   accidental
coincidences and secondary scatterings in the detector material as  well  as
positron  thermalization  process  in   matter,   different   lifetimes   of
orthopositronium in  different  materials,  momentum  distributions  due  to
quantum electrodynamic (QED) effects as well as  efficiency  for  the  gamma
quanta detection. Detailed description of these effects can  be  found  e.g.
in~\cite{Kowalski:2015bua,gonzalez,Berestetskii}.

\begin{figure}[h!]
    \centering
    \includegraphics[width=0.7\textwidth]{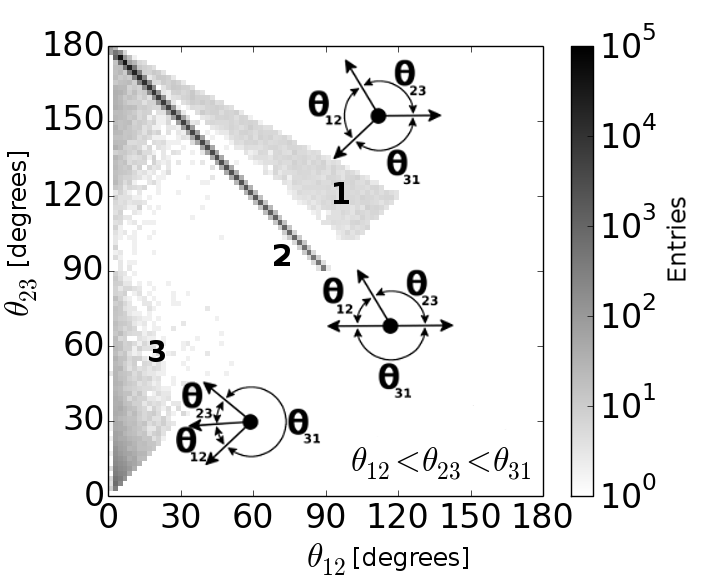}
    \caption{Distribution of relative angles between reconstructed directions of gamma quanta.
    The numbering of quanta was assigned such that $\theta_{12} < \theta_{23} < \theta_{31}$.
     Shown distributions were obtained requiring three hits each with energy deposition larger than 50 keV.
     Typical topology of $\mbox{o-Ps} \rightarrow 3\gamma$ (region 1)
     and two kinds of background events (regions 2 and 3 from $2\gamma$ events with 
     the secondary scattering in the detector) is indicted.}
    \label{fig_1}
\end{figure}
Main source of  background  contains  events  from  direct  annihilation  or
parapositronium decay where one of gamma scattered  and  was  registered  by
the detector. However, those events can be rejected by requiring small  time
differences between registration of gamma quanta, because  scattered  events
need extra time to travel to the other  part  of  the  detector.  Similarly,
after ordering the relative angles ($\theta_{12} < \theta_{23} < \theta_{31}$)  the  true  and  false
events have very small overlap region at the $\theta_{23}$ vs  $\theta_{12}$  correlation  plot
(Figure~\ref{fig_1}).
For selected events, a novel reconstruction algorithm (analogous to the  one
described  in  Ref.~\cite{Gajos:2013nra,Gajos:2015qsa})  allows  to  obtain  the   time   and   spatial
coordinates of the orthopositronium decay point by  using information  about
time of interaction of gamma quanta in the  detector~\cite{Gajos:2013nra}.  The  information
available for $i$-th hit includes its spatial  location  and  recording  time.
The problem of localizing the vertex is, in its principle,  similar  to  GPS
positioning and can be solved in a similar manner.

We expect that J-PET detector should allow for a significant improvement  in
sensitivity for  $\mathcal{CP}$  and  $\mathcal{CPT}$  tests  with  respect  to  the  best  previous
experiments~\cite{Yamazaki:2009hp,vetter:2003}. The appraisal is based on preliminary  simulations  which
will  be  described  in  details  in  the  forthcoming   publications.   The
improvement is expected mainly because of  about  two  orders  of  magnitude
larger statistics, which can be achieved due to the  possibility  of  longer
runs (within next three years in total about one  year  of  data  taking  is
planned) and due to the usage of the higher activity of positron source  (10
MBq at J-PET vs. 1 MBq at~\cite{Yamazaki:2009hp} or 0.04 MBq at~\cite{vetter:2003}). The rate  limitations  of
previous experiments are overcome by J-PET detector due to its  much  higher
granularity and about one to two orders of  magnitude  shorter  duration  of
signals (plastic scintillators at J-PET~\cite{Moskal:2014rja,Moskal:2014sra} vs. LYSO~\cite{Yamazaki:2009hp}
 or  HPGe/BGO~\cite{vetter:2003})
leading to the significant reduction of pile-ups.  In addition, it  is  also
important to stress that the J-PET detector  is  characterized  by  about  3
times higher angular resolution and most importantly that   the  J-PET  time
resolution ($\sim0.1$ns)~\cite{Moskal:2014rja,Moskal:2014sra} is improved by about a factor of ten with  respect
to experiment~\cite{Yamazaki:2009hp},  and by about a factor  of  fifty  with  respect  to  the
Gammasphere  detector~\cite{vetter:2003}.

\section*{Acknowledgments}

We acknowledge technical and administrative  support  of  T.  Gucwa-Rys,  A.
Heczko, M. Kajetanowicz, G. Konopka-Cupial, W.  Migdal,  and  the  financial
support by the Polish National Center for Development and  Research  through
grant INNOTECH-K1/IN1/64/159174/NCBR/12, the Foundation for  Polish  Science
through MPD programme, the EU  and  MSHE  Grant  No.  POIG.02.03.00-161  00-
013/09, Doctus  -  the  Lesser  Poland  PhD  Scholarship  Fund,  and  Marian
Smoluchowski Krakow Research Consortium ”Matter-Energy-Future”.


\end{document}